\documentclass[aps,prd, nofootinbib,superscriptaddress, article,8pt]{revtex4}

\usepackage{amsmath, amsfonts, amsthm, amssymb, graphicx, slashed}
\usepackage{subfig}
\usepackage{epstopdf}

\def\be{\begin{equation}}
\def\ee{\end{equation}}
\def\ba{\begin{eqnarray}}
\def\ea{\end{eqnarray}}

\def\bdm{\begin{displaymath}}
\def\edm{\end{displaymath}}

\def\bq{\begin{quote}}
\def\eq{\end{quote}}

\def\ltap{\ \raise.3ex\hbox{$<$\kern-.75em\lower1ex\hbox{$\sim$}}\ }
\def\gtap{\ \raise.3ex\hbox{$>$\kern-.75em\lower1ex\hbox{$\sim$}}\ }
\def\gl{\ \raise.5ex\hbox{$>$}\kern-.8em\lower.5ex\hbox{$<$}\ }
\def\roughly#1{\raise.3ex\hbox{$#1$\kern-.75em\lower1ex\hbox{$\sim$}}}

 at 10truept

\newcommand{\beq}{\begin{equation}}
\newcommand{\eeq}{\end{equation}}
\newcommand{\bea}{\begin{eqnarray}}
\newcommand{\eea}{\end{eqnarray}}
\newcommand{\beqa}{\begin{eqnarray}}
\newcommand{\eeqa}{\end{eqnarray}}

\begin{document}
\title{ Spherically symmetric solutions of light  Galileon }

\author         {D. Momeni}
\affiliation    {Eurasian International Center for Theoretical Physics and Department of General \& Theoretical Physics, Eurasian National University, \\ Astana 010008, Kazakhstan}
\author        {M. J. S. Houndjo}
\affiliation { Institut de Math\'ematiques et de Sciences Physiques (IMSP) - 01 BP 613 Porto-Novo, B\'enin;\\ Facult\'e des Sciences et Techniques de Natitingou, Universit\'e de Parakou, Bénin.}
\author        {E. G\"{u}dekli}
\affiliation { Department of Physics, Istanbul University, Istanbul,
Turkey.}

\author {M. E. Rodrigues}
\affiliation{ Faculdade de Ci\^encias Exatas e Tecnologia, Universida
de Federal do Par\'a \\ Campus Universit\'ario de Abaetetuba, CEP 68440-000, Abaetetuba, Par\'a, Brazil}

\author {F. G. Alvarenga}
\affiliation{Departamento de Engenharia e Ci\^encias Naturais - CEUNES \\ Universidade
Federal do Esp\'ırito Santo - CEP 29933-415 - S\~ao Mateus/ ES,
Brazil.}

\author         {R. Myrzakulov}
\affiliation    {Eurasian International Center for Theoretical Physics and Department of General \& Theoretical Physics, Eurasian National University, \\ Astana 010008, Kazakhstan}
\date{\today}
 
\begin{abstract}
We have been studied the  model of light Galileon with translational shift symmetry $\phi\to \phi+c$. The  matter Lagrangian is presented  in the form $\mathcal{L}_{\phi}= -\eta (\partial \phi)^2+\beta G^{\mu\nu}\partial_{\mu}\phi\partial_{\nu}\phi$. We have been addressed two issues: the first is that, we have been proven that, this type of  Galileons belong to the modified matter-curvature models of gravity in type of $f(R,R^{\mu\nu}T_{\mu\nu}^m)$. Secondly, we have been investigated exact solution for spherically symmetric geometries in this model. We have been found an  exact solution  with singularity at $r=0$ in null coordinates. We have been proven that the solution has  also a non-divergence current vector norm. This solution can be considered as an special solution which has been investigated in literature before, in which the Galileon's field is non-static (time dependence). Our scalar-shift symmetrized Galileon has the simple form of $\phi=t$, which it is remembered by us dilaton field. 
\end{abstract}

\maketitle
\section{Introduction}
It is well established that Einstein gravity is the best description gravity as a gauge theory \cite{Moshe}. It predicts relativistic effects like light's bending and the existence of most compact objects of the Universe, black holes perfectly. To have a more precise description of our Universe we need to be able to explain not only our solar system but the large scales, in cosmological scales. When we observed our Universe evolve in an accelerated sheme\cite{obs1}-\cite{obs3}, and most part of the matter (energy) of our Universe is unknown, we fell to work with modified gravities, one of the best candidates to have a reasonable extension of Einstein gravity to large scales (see \cite{Nojiri:2010wj}-\cite{ijgmmp}). There are several types of modified gravities. One of the first ones was $f (R) $ gravity \cite{Buchdahl} in which we just replace the usual scalar term $R$ by an arbitrary function $f (R) $ which it was found many applications in cosmology. Another approach is to address gravity as gauge theory but from a different point of view, in which the space-time does not have any curvature but it has "torsion", $T$, it has been introduced as a valid modification (different description) of gravity \cite{Bengochea:2008gz}-\cite{Myrzakulov:2010vz} and it has been investigated to resolve several problems of cosmology \cite{Jamil:2012nma} -\cite{Rodrigues:2013ifa}.
 
One particularly interesting view is to keep Einstein gravity and add some types of scalar scalar-tensor Lagrangians . The main point is to keep the order of the equation of motion at second order and not higher, to avoid problems with higher derivatives like Ostrogradski instability \cite{ostro}. The most general form of such scalar-tensor Lagrangians was founded by
Horndeski \cite{horndeski:1974} and it has recently been rediscovered independently in \cite{general}). Such higher order models of scalar-tensor is naturally appeared in Brans-Dicke gravity \cite{bdgravity} or some recently studied models like \cite{covgal,galmodels}. The name of a scalar field is changed to gallon since the model respects to the Galilean symmetry $\phi\to \phi+b_{\mu} x^{\mu}+c$, and the models are called by Galilean theory \cite{galileon}. One station to merit such models as higher order scalar-tensor Lagrangians but with second order field equations, is when we are investigating the Kaluza-Klein compactifications scheme in the framework of Lovelock densities in higher dimensional spacetimes\cite{kkl, VanAcoleyen:2011mj}. Recently much attentions have been done of galileons \cite{Kobayashi:2011nu}-\cite{Charmousis:2011ea} as one desirable model of scalar field of early and late time history of Universe. \par
In this paper we study the type of Galileons with translational symmetry $\phi\to\phi+c$. The simplest lower order Lagrangian of such models is contained by two terms $(\partial\phi)^2,G^{\mu\nu}\partial_{\mu}\phi\partial_{\nu}\phi$. The Lagrangian is written as the following $\mathcal{L}_m= -\eta (\partial \phi)^2+\beta G^{\mu\nu}\partial_{\mu}\phi\partial_{\nu}\phi$ where $\eta,\beta$ are a couple of parameters. We address two issues about such models:

\begin{itemize}
\item The model is identified as a modified gravity in the form of  $f(R,R^{\mu\nu}T_{\mu\nu}^m)$.
\item The model posses the exact spherically symmetric  solution with time dependent galileon. 
\end{itemize}
Before in literature it was proven that such solution exists  \cite{Babichev:2013cya} and has non-divergence norm, in our work we also address this important issue about regular solutions. Our plan in this letter is as the following: In Sec. (\ref{Horndeski}) we review Horndeski's models from a modern point of view.  In Sec. (\ref{equiv}) we will prove the formal equivalence between this type of Galileon and matter-curvature coupled modified gravity. In Sec. (\ref{exact}) we will find the exact solution in spherical symmetry. In Sec. (\ref{per}) we will study perturbations. The last section is devoted to conclude the results.

\section{Horndeski's  model of scalar-tensor gravity}\label{Horndeski}
As we said it before, the idea of the construction of scalar-tensor models with the most general higher order terms but with second order field equation was belonged to \cite{horndeski:1974} who proposed the Lagrangian densities from a mathematical point of view . But very recently the Horndeski's densities have been revisited in an elegant new form in \cite{Charmousis:2011bf} (see \cite{Charmousis:2011ea} for associated equations of motion). To revisit Horndeski's model in new dress, let us start with the following general second-order scalar tensor theory is given by: 
\be 
\label{eq:action}
S=S_{H}[g_{\mu\nu}, \phi]+S_m[g_{\mu\nu}; \Psi_n]
\ee
Here the Horndeski action, $S_ {H} = \int d^4 x \sqrt{-g} {\cal L} _H$, can be rewritten from the equation (4.21) of \cite{horndeski:1974}, in the following non-trivial form:
 \ba
\label{eq:horndeskiFullLag}
{\cal L}_H&=& \kappa_1(\phi ,\rho)\delta^{\alpha \beta \gamma} _{\mu\nu\sigma}\nabla^\mu\nabla_\alpha \phi  R_{\beta \gamma} ^{\;\;\;\;\nu\sigma}
           -\frac{4}{3}\kappa_{1,\rho}(\phi ,\rho)\delta^{\alpha \beta \gamma} _{\mu\nu\sigma}\nabla^\mu\nabla_\alpha \phi\nabla^\nu\nabla_\beta \phi\nabla^\sigma\nabla_\gamma \phi \\\nonumber
        &~&+\kappa_3(\phi ,\rho)\delta^{\alpha \beta \gamma} _{\mu\nu\sigma}\nabla_\alpha \phi \nabla^\mu\phi  R_{\beta \gamma} ^{\;\;\;\;\nu\sigma}
           -4\kappa_{3,\rho}(\phi ,\rho)\delta^{\alpha \beta \gamma} _{\mu\nu\sigma}\nabla_\alpha \phi \nabla^\mu\phi \nabla^\nu\nabla_\beta \phi \nabla^\sigma\nabla_\gamma \phi \\\nonumber
        &~&+[F(\phi ,\rho)+2W(\phi )]\delta_{\mu\nu}^{\alpha \beta }R_{\alpha \beta }^{\;\;\;\;\mu\nu}
           -4F(\phi,\rho)_{,\rho}\delta_{\mu\nu}^{\alpha \beta }\nabla_\alpha \phi\nabla^\mu\phi \nabla^\nu\nabla_\beta \phi \\\nonumber
        &~&-3[2F(\phi ,\rho)_{,\phi }+4W(\phi )_{,\phi }+\rho\kappa_8(\phi,\rho)]\nabla_\mu\nabla^\mu\phi 
           +2\kappa_8\delta_{\mu\nu}^{\alpha \beta }\nabla_\alpha \phi \nabla^\mu\phi \nabla^\nu\nabla_\beta \phi \\\nonumber
        &~&+\kappa_9(\phi ,\rho),\\\nonumber
\rho&=&\nabla_\mu\phi \nabla^\mu\phi ,
\ea
where $\{\kappa_i(\phi,\rho)\}_{i=1,3,8,9}$, are a set of arbitrary functions of the scalar field (Galileon) $\phi$ . To have dynamic for Galileon, we add kinetic part of $\rho$ . Furthermore, we know that:
\ba
\label{eq:Fdef}
F_{,\rho}&=&\kappa_{1,\phi}-\kappa_3-2\rho\kappa_{3,\rho}
\ea
here $W(\phi)$ is an arbitrary function of the scalar field. The most important comment which we must indicate here is that {\it Horndeski's theory is equivalent to the generalised scalar tensor theory} if we limited ourself to four dimensions \cite{general}-\cite{VanAcoleyen:2011mj}. The main reference about this equivalence is \cite{Kobayashi:2011nu}. This last reference also provided the full catalog to work with Horndeski model. Except then the model given by (\ref{eq:horndeskiFullLag}) is too much complicated, a suitable simple Lagrangian's form can be derived in which it respects to the translational(shift) symmetry of field $\phi\to\phi+c$, which it is written in the most simplest form as the following:
\bea
\mathcal{L}=-\eta (\partial \phi)^2+\beta G^{\mu\nu}\partial_{\mu}\phi\partial_{\nu}\phi
\eea
 In next section we shall show that how this minimal Horndeski's model is equivalent to the modified gravity in the form of curvature-matter models.

\section{Equivalence of light Galileon to curvature-matter coupled models of gravity}\label{equiv}
A class of modified theories of gravity has been discovered in which geometry is coupled to the matter field(s) non minimally, via Lagrangian of matter or energy-momentum tensor , with the following general Lagrangian \cite{Harko:2014gwa}:
\bea
S=\int {d^4 x \sqrt{-g}f(R,T_{\mu}^{\mu},R^{\mu\nu}T_{\mu\nu},\mathcal{L}_m)}
\eea
Where $\mathcal{L}_m=-\eta(\partial\phi)^2$, $T_{\mu\nu}^m=\mathcal{L}_m g_{\mu\nu}-2\frac{\partial\mathcal{L}_m}{\partial g^{\mu\nu}}$. For the simple case of $f(R,T_{\mu}^{\mu}) $it has been investigated in several problems from cosmology to black holes \cite{Harko:2014gwa} -\cite{Alvarenga:2013syu}. The new extension of such models has been proposed recently as \cite{Odintsov:2013iba},\cite{Haghani:2013oma}
$f(R,R^{\mu\nu}T_{\mu\nu}^m)$. Now we prove a theorem in which we show that Galileon  with translational symmetry is equivalent to a type of $f(R,R^{\mu\nu}T_{\mu\nu}^m)$ theory.\\
\\
\texttt{Theorem}: {\it 
Translational invariance Galileon lagrangian with the following action}
:
\begin{eqnarray}
&&S=\int {d^4 x \sqrt{-g}\Big(\xi R -\eta (\partial \phi)^2+\beta G^{\mu\nu}\partial_{\mu}\phi\partial_{\nu}\phi\Big)}\label{S1}.
\end{eqnarray}
{\it is equivalent to the } $f(R,R^{\mu\nu}T_{\mu\nu}^m)$ theory.
\\
\\
\texttt{Proof}:
{\it Firstly we should clarify that the action given by $S_{\phi}$ is just the translational summarized galileon model. 
To have the correct form of the Einstein-Hilbert action as the limiting form, we set $\xi=\frac{1}{16\pi G} $. The notations are $(\partial \phi)^2=g^{\mu\nu}\partial_{\mu}\phi\partial_{\nu}\phi$. $G^{\mu\nu}$ is Einstein tensor is defined by $G^{\mu\nu}=R^{\mu\nu}-\frac{R}{2}g^{\mu\nu}$. The model is invariant under translational symmetries $\phi\to\phi+c$ \cite{Babichev:2013cya}. Phenomenologically, we can say that such models describe light Galileon, the Galileons without potential forms. We rewrite we introduce the Lagrangian of matter fields as the following:
\bea
\mathcal{L}_m=-\eta(\partial \phi)^2
\eea
The associated energy-momentum tensor is written as the following:
\bea
T_{\mu\nu}^m=\eta\Big[2\partial_{\mu}\phi\partial_{\nu}\phi-g_{\mu\nu}(\partial \phi)^2\Big]
\eea
we can write the following expression:
\bea
R^{\mu\nu}T_{\mu\nu}^m=\eta\Big[2R^{\mu\nu}\partial_{\mu}\phi\partial_{\nu}\phi-R(\partial \phi)^2\Big]
\eea
Using $G^{\mu\nu}=R^{\mu\nu}-\frac{R}{2}g^{\mu\nu}$ we rewrite it as the following:
\bea
R^{\mu\nu}T_{\mu\nu}^m=2\eta G^{\mu\nu}\partial_{\mu}\phi\partial_{\nu}\phi
\eea
consequently it is possible to write
(\ref{S1}) in the following equivalent form:
\begin{eqnarray}
&&S=\int {d^4 x \sqrt{-g}\Big(\xi R+\frac{\beta}{2\eta}R^{\mu\nu}T_{\mu\nu}^m\Big)}+\int{d^4x \sqrt{-g}\mathcal{L}_m}\label{S2}.
\end{eqnarray}
 where $\mathcal{L}_m=-\eta  (\partial \phi)^2 $, so now we observe that the model is in the following modified gravities  models.}
{\bf Q.E.D}
\\
\\

To find the equation of motion (EOM) of metric field , we vary the action given by Eq.~(\ref{S2}) with respect to the metric , we obtain:
\begin{eqnarray}
&&f_{R}G_{\mu\nu}+g_{\mu\nu}\Big[\Box f_{R}+\frac{R}{2}f_{R}-\frac{f}{2}\Big]-\nabla_{\mu}\nabla_{\nu}f_{R}-\frac{T_{\mu\nu}}{2\xi}=0.
\end{eqnarray}
For our model we've:
\begin{eqnarray}
&&G_{\mu\nu}-\frac{\beta}{2\xi}g_{\mu\nu}\Big[G^{\alpha\beta}\partial_{\alpha}\phi\partial_{\beta}\phi\Big]=\frac{\eta}{2\xi}\Big(2\partial_{\mu}\phi\partial_{\nu}\phi-g_{\mu\nu}(\partial\phi)^2\Big).
\end{eqnarray}
Or equivalently:
\begin{eqnarray}
&&R_{\mu\nu}+\frac{\beta}{2}\Big[G^{\alpha\beta}\partial_{\alpha}\phi\partial_{\beta}\phi\Big]g_{\mu\nu}
=-\mu\partial_{\mu}\phi\partial_{\nu}\phi,\label{eom1}.
\end{eqnarray}
here $\mu=-\frac{\eta}{\xi}$. 
We mention here that we have also the following EOM for $\phi$ (or the momentum constraint ) as the generalized Klein-Gordon equation:
\begin{eqnarray}
&&\nabla_{\nu}\Big[\beta G^{\mu\nu}\nabla_{\mu}\phi-\eta\nabla^{\nu}\phi\Big]=0.
\end{eqnarray}
Using the metricity condition $\nabla_{\nu}g^{\mu\nu}=0$ and Bianchi's identity $\nabla_{\nu}G^{\mu\nu}=0$ we are able to rewrite the above EOM is the following form:
\begin{eqnarray}
&&\Big[\beta G^{\mu\nu}-\eta g^{\mu\nu}\Big]\nabla_{\nu}\nabla_{\mu}\phi=0\label{eom2}.
\end{eqnarray}
It is adequate to interpret $\beta G^{\mu\nu}-\eta g^{\mu\nu}$ as the induced metric of $h^{\mu\nu}=\beta G^{\mu\nu}-\eta g^{\mu\nu}$. So, (\ref{eom2}) reads as the following:
\bea
h^{\mu\nu}\nabla_{\nu}\nabla_{\mu}\phi=0.
\eea
In the next section we are looking for exact spherically symmetric solutions of (\ref{eom1},\ref{eom2}).

\section{Searching for exact spherically symmetric solutions}\label{exact}
 Our aim here is to find an exact solution for static spherically symmetric metric in the following form:
\begin{eqnarray}
&&ds^2=A(r)dt^2-B(r)dr^2-r^2d\Omega^2\label{g}.
\end{eqnarray}
here $d\Omega^2\equiv d\theta^2+\sin^2\theta d\varphi^2$. 
The form of metric given by $A(r)B(r)=1$,  is not required because we are not sure that {\it null energy condition satisfies  as well as the fact that the radial photon experiences  acceleration or not} \cite{Dadhich:2012pda}.\par
For metric given by (\ref{g}) we rewrite (\ref{eom2}) in the following form:
\begin{eqnarray}
&&\Big[\beta G^{tt}-\eta g^{tt}\Big]\Big(\ddot{\phi}-\frac{A'}{2B}\phi'\Big)+\Big[\beta G^{rr}-\eta g^{rr}\Big]\Big(\phi''-\frac{B'}{2B}\phi'\Big)=0.
\end{eqnarray}
Note that $\phi=\phi(t,r)$. In general this equation is too much complicated to solve. Here we are interesting to study non-static (time dependent) Galileon (for a more general case see \cite{Babichev:2013cya}). By this reason, we assume that $\phi=\phi(t)$. By this Ansatz we obtain:
\begin{eqnarray}
&&\Big[\beta G^{tt}-\eta g^{tt}\Big]\ddot{\phi}=0.
\end{eqnarray}
One possibility is to have $\ddot{\phi}=0$ but $\Big[\beta G^{tt}-\eta g^{tt}\Big]\neq0$. If we relax $\ddot{\phi}=0$, the case with $\Big[\beta G^{tt}-\eta g^{tt}\Big]\neq 0$ and $\Big[\beta G^{rr}-\eta g^{rr}\Big]=0, \ddot{\phi}\neq0$ is well established in \cite{Babichev:2013cya}. Also in that Ref.\cite{Babichev:2013cya}, the authors consider $\phi=\phi(t,r)$. Also they imposed that $J^ {r} =0$ because they required to have regular solutions in which the norm of current vector remains finite on horizon. Actually they found a rich family of solutions by the above assumptions.  
In our letter, we explore for exact solutions in which $J^{r}\neq 0,\ \ \ddot{\phi}=0,\ \ J^{t}=0$. 
So, we assume that meanwhile:
\begin{eqnarray}
&&\Big[\beta G^{tt}-\eta g^{tt}\Big]=0 \ \ \&\& \label{eq2}\ \ \ddot{\phi}=0.
\end{eqnarray}
With this constraint,the momentum conservation (\ref{eom2}) is satisfied identically. So, using this constraint we obtain:
\begin{eqnarray}
&&G^{\alpha\beta}\partial_{\alpha}\phi\partial_{\beta}\phi
=G^{tt}\dot{\phi}^2=\frac{\eta}{\beta} g^{tt}.
\end{eqnarray}
Here $\dot{\phi}=1$. Consequently, the $tt$,$rr$ components of the (\ref{eom1}) read:
\begin{eqnarray}
&&R_{tt}=-\Big(\mu+\frac{\eta}{2}\Big),\ \ 
R_{rr}=\frac{\eta}{2}\frac{B}{A}.
\end{eqnarray}
Using (\ref{g}) we obtain:
\begin{eqnarray}
&&\frac{1}{2}\frac{A''}{B}-\frac{1}{4}\frac{A'^2}{AB}-\frac{1}{4}\frac{A'B'}{B^2}+\frac{A'}{rB}=\mu+\frac{\eta}{2},
\\&&
\frac{1}{2}\frac{A''}{B}-\frac{1}{4}\frac{A'^2}{AB}-\frac{1}{4}\frac{A'B'}{B^2}-\frac{A'}{rB}=\frac{\eta}{2}.
\end{eqnarray}
By subtraction we find:
\begin{eqnarray}
&&\frac{2A'}{rB}=\mu\label{1-2}.
\end{eqnarray}
So, the general form of the meric is written as the following:
\begin{eqnarray}
&&ds^2=A(r)dt^2-\frac{2A'}{\mu r}dr^2-r^2d\Omega^2.
\end{eqnarray}
To find metric function $A(r)$, we substitue (\ref{1-2}) in one of the equations  to obtain:
\begin{eqnarray}
&&\frac{A''}{A'}-\frac{A'}{A}=\frac{4\eta+3\mu}{\mu}\frac{1}{r}.
\end{eqnarray}
We solve it to find:
\begin{eqnarray}
A(r)=A_0 e^{\frac{c}{n+1}r^{n+1}},\label{A}\ \ n=\frac{4\eta+3\mu}{\mu}.
\end{eqnarray}
Finally we present one exact solution for the system as the following:
\begin{eqnarray}
&&ds^2=A_0 e^{\frac{c}{n+1}r^{n+1}}\Big(dt^2-\frac{2c}{\mu }r^{n-1}dr^2\Big)-r^2d\Omega^2.
\end{eqnarray}
Or equivalently :
\begin{eqnarray}
&&ds^2=A_0 e^{\frac{c}{4(1-\xi)}r^{4(1-\xi)}}\Big(dt^2+\frac{2c\xi r^{2(1-2\xi)}}{\eta }dr^2\Big)-r^2d\Omega^2,\\&&
\phi(t;r)=t.
\end{eqnarray}
In the units which $8\pi G=1$, $\xi=\frac{1}{2}$, so the metric reads:
\begin{eqnarray}
&&ds^2=A_0 e^{-\frac{|c|}{2}r^{2}}\Big(dt^2-\frac{|c|}{\eta }dr^2\Big)- r^2d\Omega^2,\label{metric}\\&&
\phi(t;r)=t.
\end{eqnarray}
Here $c0$. Since $A_0$ to be considered as an arbitrary constant, we chose it as $A_0=\frac{\eta}{|c|}$. In this case, we are able to define a new time coordinate $\tau=\sqrt{\frac{\eta}{|c|}}t$ and a pair of null coordinates $u=\tau-r,v=\tau+r$, in which the metric is written as the following:
\begin{eqnarray}
&&ds^2= e^{-\frac{|c|}{8}(v-u)^{2}}dudv- \frac{(v-u)^2}{4}d\Omega^2,\label{metric2}\\&&
\phi(v,u)=\sqrt{\frac{|c|}{\eta}}\frac{v+u}{2}.
\end{eqnarray}
The Kretchmann invariant $\mathcal{K}=R^{\alpha\beta\gamma\delta}R_{\alpha\beta\gamma\delta}$ of the metric is given by:
\begin{eqnarray}
&&\mathcal{K}={\frac {Y_1-128\,{
{\rm e}^{-1/8\, \left(  \left| c \right|  \right) ^{2} \left( -v+u
 \right) ^{2}}}+5\, \left(  \left| c \right|  \right) ^{4}{v}^{4}+64}{
 \left( {{\rm e}^{-1/8\, \left(  \left| c \right|  \right) ^{2}
 \left( -v+u \right) ^{2}}} \right) ^{2} \left( -v+u \right) ^{4}}},\\&&\nonumber
Y_1=-20\, \left(  \left| c \right|  \right) ^{4}{v}^{3}u+30\,
 \left(  \left| c \right|  \right) ^{4}{v}^{2}{u}^{2}-20\, \left( 
 \left| c \right|  \right) ^{4}v{u}^{3}+5\, \left(  \left| c \right| 
 \right) ^{4}{u}^{4}+64\, \left( {{\rm e}^{-1/8\, \left(  \left| c
 \right|  \right) ^{2} \left( -v+u \right) ^{2}}} \right) ^{2}
\end{eqnarray}
So, as we observe,  there is a singularity located at $u=v$ or $r = 0$ ), as the Schwarzschild solution in general relativity.

We should check $J^{r}$ for this exact solution. We have:
\begin{eqnarray}
J^{r}=\frac{\beta}{B^2}\Big(\frac{B-1}{r^2}-\frac{A'}{rA}\Big)+\frac{\eta}{B}
\end{eqnarray}
Using (\ref{1-2}) and  (\ref{A}) we obtain  $J^r\neq 0$. We conclude:
\begin{eqnarray}
J^2=J_{\mu}J^{\mu}=-\frac{J_{r}^2}{B}\neq0,\ \ J^t\equiv 0, \ddot{\phi}=0.
\end{eqnarray}

\section{Slightly modified Xanthopolous and  Zannias solutions}\label{per}
If $\beta=0$, there exits an exact solution   for (\ref{eom1}) was obtained \cite{Buchdahl:1959nk}. 
Buchdahl's solution as the following:
\begin{eqnarray}
&&ds^2=(1-\frac{2m}{r})^{\zeta}dt^2-(1-\frac{2m}{r})^{-\zeta}dr^2-r^2(1-\frac{2m}{r})^{1-\zeta}d\Omega^2.
\end{eqnarray}
Where $\phi^0=\lambda\ln(1-\frac{2m}{r}),\ \ \zeta=\pm\sqrt{1-2\mu\lambda^2}$. It has been proven that the above metric with scalar field  $\phi\equiv \phi^0$ is an exact solution of Einsten field equations with massless scalar field with the following action:
\begin{eqnarray}
S=\int{d^4x \sqrt{-g}\Big(R+\mu\nabla_{\alpha}\phi\nabla^{\alpha}\phi\Big)}.
\end{eqnarray}
Which it satisfies the equation of motion $R_{\alpha\beta}=-\mu\nabla_{\alpha}\phi\nabla_{\beta}\phi$. The solution was founded by Buchdahl by an intuitive and elegant application of reciprocal metrics. It was proven that the solution is a black hole with a singularity at $r=0$ and the horizon located at $r=2m$. Furthermore, it was shown that 
in the absence of scalar field, $\mu=0$, the solution is recovred by the empty spacetime  Schwarzschild  solution. Our aim here is to extend and use Buchdahl's solution to find an approximated solution for (\ref{eom1}). We take in to the account that the nonlinearity parameter $\beta$ is smallness parameter and we try to extend the solutions of (\ref{eom1}) in a series of perturbations.  Another form of the two parameters family of exact solutions founded later for a general $D\geq4$ spacetime, which is in the following form(is written for $D=4$ \cite{Xanthopolous}
\footnote{We changed the singnature to kept the regularity of our work}:
\begin{eqnarray}
ds^2=\Big[\frac{r-r_{0}}{r+r_{0}}\Big]^{2\gamma}dt^2-\Big[1-(\frac{r_0}{r})^2\Big]^2\Big[\frac{r-r_{0}}{r+r_{0}}\Big]^{-2\gamma}\Big(dr^2+r^2d\Omega^2\Big)\label{metric1989}.
\end{eqnarray}
Where the scalar field is obtained  by the following form:
\begin{eqnarray}
\phi(r)=\sqrt{2(1-\gamma^2)}\ln\Big[\frac{r-r_{0}}{r+r_{0}}\Big]\label{phi1989}
\end{eqnarray}
As it was shown , the spacetime is asyptotically flat, furthermore the scalar field vanishes at infinity $r\to \infty$. In the absence of scalar field, when $\gamma=1$, the spacetime reduces to the Schwarzschild  metric which is written in isotropic coordinates. The case with $\gamma=-1$ represents the Schwarzschild solution with negative mass  which it has  a naked singularity.  

\par
Our aim here to find the exact solution (black hole) with $\beta\neq0$ for (\ref{eom1}). One possibility is to use perturbation method around this exact solution ($\beta=0$ case) as the following:
\begin{eqnarray}
&&g_{\mu\nu}=g_{\mu\nu}^{0}+\beta\delta g_{\mu\nu}+...,\\
&&\phi=\phi^{0}(r)+\beta\delta\phi(t,r)+...
\end{eqnarray}
where $\{g_{\mu\nu}^{0},\phi^{0}(r)\}$   are given by (\ref{metric1989},\ref{phi1989})  and try to find the first order corrections $\{\delta g_{\mu\nu},\delta\phi(t,r)\}$. 
The perturbated equation in first order $\mathcal{O}(\beta)$ is written as the following:
\begin{eqnarray}
&&\delta R_{\mu\nu}+\frac{1}{2}(\partial_{r}(\phi^0))^2(g_{\mu\nu}^0)\delta G^{rr}=-\mu\Big[\partial_{\mu}\phi^0\partial_{\nu}\delta\phi
+\partial_{\nu}\phi^0\partial_{\mu}\delta\phi\Big].
\end{eqnarray}
Here
\begin{eqnarray}
&&\delta R_{\mu\nu}=\nabla_{\rho}(\delta\Gamma^{\rho}_{\mu\nu})-\nabla_{\nu}(\delta\Gamma^{\rho}_{\rho\mu})
\end{eqnarray}
where
\be
\delta\Gamma^{\sigma}_{\mu\nu}=\frac{1}{2}g^{\sigma\lambda}\left(\delta g_{\mu\lambda;\nu}+\delta g_{\nu\lambda;\mu}-\delta g_{\mu\nu;\lambda}\right)\ .
\label{I4c}
\ee
We assume that the perturbated metric is given by :
\begin{eqnarray}
ds^2=\Big[\frac{r-r_{0}}{r+r_{0}}\Big]^{2\gamma}e^{\beta F(r)}dt^2-\Big[1-(\frac{r_0}{r})^2\Big]^2\Big[\frac{r-r_{0}}{r+r_{0}}\Big]^{-2\gamma}e^{-\beta H(r)}\Big(dr^2+r^2d\Omega^2\Big)\label{perg}.
\end{eqnarray}
where the perturbations functions $\{F(r),H(r)\}$ to be small in comparison to the $g_{\mu\nu}^0$. If we compute the Einstein tensor for (\ref{perg}) we obtain, in the first order perturbation theory of $\{\mathcal{O}(F(r)),\mathcal{O}(H(r))\}$, the component $\delta G^{rr}$ is given by :
\bea
\delta G^{rr}=-\Big(( {r}^{4}-2\,{r}^{3}r_0\gamma+2\,r{r_{0}}^{3}\gamma-{r_{0}}^{4}
) F '( r) + ( -{r}^{4}+{r_{0}}^{4}
) H' ( r )\Big) {r}^{7}( r-r_{0}) ^{-6+4\,\gamma}
( r+r_{0}) ^{-6-4\,\gamma}
\eea
Also we read:
\begin{eqnarray}
&&\delta R_{tt}= -{\frac {{r}^{4}( {r}^{2}F'' +2rF'+2 F' r_{0}\gamma-2\gamma r_{0}H'
 - F'' {r_{0}}^{2}) }{2( r+r_{0}) ^{3}( r-
r_{0}) ^{3}}}
 \\&&
\delta R_{rr}= \frac{1}{2r(r^2-r_0^2)}\Big(( -2H''  +F '') {r}^{3}-2H'   {r}^{2}+ ( 2\gamma r_{0}H' - F'' r_{0}^{2}+6F' r_{0}\gamma+2H''
    r_{0}^{2} ) r\\&&\nonumber+2H'
 r_{0}^{2}-2( F'
 ) r_{0}^{2}\Big)
\end{eqnarray}
So, we obtain:
\begin{eqnarray}
&&tt:\ \ ( -2r\gamma r_{0}+{r}^{2}+r_{0}^{2} )F'
 - H' 
 ( r_{0}^{2}+{r}^{2}) + ( {r}^{2}-r_{0}^{2}) {
}F''  + ( 2\gamma r_{0}+2r
)F'  -2\gamma r_{0}H'=0,\\&&
rt:\ \ \partial_t(\delta\phi(t,r))=0\\&&
rr:\ \  4( \gamma^2-1)( ( -
2r_0r\gamma+r_0^{2}+{r}^{2}) F' 
-H'  (r_0^{2}+{r
}^{2}) ) {r}^{3}( \frac {r-r_0}{r+r_0}) ^{2
\gamma}r_0^{2}+r ( r-r_0) ^{2}( r+r_0) ^{2}
F'' \\&&\nonumber-2\,r ( r-r_0) ^{
2}( r+r_0) ^{2}H''
+4r_0\partial_r (\delta\phi( t,r)) -2( 
( -3r_0r\gamma+r_0^{2}) F' +  H' ( -
r_0r\gamma+{r}^{2}-r_0^{2}))( r^2-r_0^2)=0.
\end{eqnarray}
An additional equation is needed which we can obtain by variation of (\ref{eom2}), it reads as the following:
\begin{eqnarray}
&&\Big[G^{\mu\nu}-\eta\delta g^{\mu\nu}\Big]\nabla_{\nu}\nabla_{\mu}\phi-\eta g^{\mu\nu}\nabla_{\nu}\nabla_{\mu}\delta\phi=0.
\end{eqnarray}
Since $\phi=\phi(r),\partial_{t}\delta\phi=0$, so we have:
\begin{eqnarray}
&&\delta\phi''-\frac{1}{2}g^{rr}\frac{\partial g_{rr}}{\partial r}\delta\phi'=0.
\end{eqnarray}
 It can be integrated to give us 
\begin{eqnarray}
\delta\phi(t,r)=C_1+C_2\int{dr \Big(1-(\frac{r_0}{r})^2\Big)\Big(\frac{r+r_0}{r-r_0}\Big)^{\gamma/2}}\label{deltaphi}.
\end{eqnarray}
If we substituing (\ref{deltaphi}) in $\{tt,rr\}$ equations we can find $\{F,H\}$.

\section{Conclusion}
 
This paper is devoted to the study of the model of the light Galileon  through the translational shift symmetry. The Galileon theory is defined to be the most general Lorentz-invariant, local model of a scalar field whose classical equation of motion possesses Galilean symmetry and presenting the avoidance of the presence of ghosts in arbitrary configurations.  In general, the Galileon theory is view as a kind of modified theory of gravity. We address here two issues to the models, first,  identifying it as a modified  
gravity in the form $f(R, R^{\mu\nu}T_{\mu\nu}^{m})$, and in the second way,  assuming that models present exact spherically solution with time dependent Galileon. We explored regular solutions and showed that they have non divergence norm, in the same way as in the literature for other type of solutions. \par
More precisely,  we  revised Horndeski's model  from point of view and showed how this minimal model is equivalent to the modified gravity in the form of curvature-matter models. By the way we point out and prove  a Theorem according to what the considered Galileon lagrangian is equivalent to the $f(R,R^{\mu\nu}T^{m}_{\mu\nu})$. View like this, the Galileon theory , with explicit  scalar field may be view as gravitational field where the matter is coupled to the gravity through the the stress tensor. This is quick clear, since the energy momentum tensor should be written depending on the  scalar field. This is an important result presented in this work.\par
On the other hand, we search for exact spherically symmetric solutions and observe that  there is singularity located at $r=0$ as it is the case for Schwartzchild solution in general relativity. Attention is also attached to type of Xanthopolous and Zannias solutions, where we  assumed that the nonlinearity parameter is very small and try to extend the solution (6) in series of perturbations and obtain the  function of perturbation depending explicitly on the radial coordinate $r$.






\end{document}